\documentclass[11pt,a4paper]{article}
\usepackage{jinstpub}
\usepackage[utf8]{inputenc}
\usepackage{csquotes}    

\title{Simulations of Future Particle Accelerators: Issues and Mitigations}
\author[a]{D.~Sagan,}
\author[b]{M.~Berz}
\author[c]{N.M.~Cook,}
\author[d]{Y.~Hao,}
\author[a]{G.~Hoffstaetter,}
\author[e]{A.~Huebl,}
\author[f]{C.-K.~Huang,}
\author[g]{M.H.~Langston,}
\author[h]{C.~E.~Mayes,}
\author[e]{C.~E.~Mitchell,}
\author[h]{C.-K.~Ng,}
\author[e]{J.~Qiang,}
\author[e]{R.~D.~Ryne,}
\author[f]{A.~Scheinker,}
\author[i]{E.~Stern,}
\author[e]{J.-L.~Vay,}
\author[j]{D.~Winklehner,}
\author[k]{H.~Zhang}

\affiliation[a]{Cornell University, Ithaca, NY, 14853, USA}
\affiliation[b]{Michigan State University, East Lansing, MI 48824}
\affiliation[c]{RadiaSoft LLC, RadiaSoft LLC, Boulder, CO 80301, USA}
\affiliation[d]{Brookhaven National Laboratory, 98 Rochester St, Upton, NY 11973, USA}
\affiliation[e]{Lawrence Berkeley National Laboratory, Berkeley, CA, 94720, USA}
\affiliation[f]{Los Alamos National Laboratory, Los Alamos, NM, 87545, USA}
\affiliation[g]{Reservoir Labs Inc., New York, NY 10012, USA}
\affiliation[h]{SLAC National Accelerator Laboratory, Menlo Park, CA, 94025, USA}
\affiliation[i]{Fermi National Accelerator Laboratory, Batavia, IL, 60510, USA}
\affiliation[j]{Massachusetts Institute of Technology, Cambridge, MA, 02138, USA}
\affiliation[k]{Thomas Jefferson National Accelerator Facility, Newport News, VA, 23606, USA}

\abstract{The ever increasing demands placed upon machine performance have resulted in the need for more comprehensive particle accelerator modeling. Computer simulations are key to the success of particle accelerators. Many aspects of particle accelerators rely on computer modeling at some point, sometimes requiring complex simulation tools and massively parallel supercomputing. Examples include the modeling of beams at extreme intensities and densities (toward the quantum degeneracy limit), and with ultra-fine control (down to the level of individual particles). In the future, adaptively tuned models might also be relied upon to provide beam measurements beyond the resolution of existing diagnostics. Much time and effort has been put into creating accelerator software tools, some of which are highly successful. However, there are also shortcomings such as the general inability of existing software to be easily modified to meet changing simulation needs. In this paper possible mitigating strategies are discussed for issues faced by the accelerator community as it endeavors to produce better and more comprehensive modeling tools. This includes lack of coordination between code developers, lack of standards to make codes portable and/or reusable, lack of documentation, among others.}

\emailAdd{david.sagan@cornell.edu}

\date{July 2021}

\begin{document}

\maketitle

\section{Introduction}

Particle accelerator simulation is critical to the design, commissioning, operation, and upgrading of accelerator facilities which cost many millions to billions of dollars. Accelerator simulation is a large, complex topic, and much time and effort has been spent in developing simulation software. Nevertheless, in the field of accelerator physics, simulation code development has often been a haphazard affair. Due to developers retiring or moving on to other projects, numerous simulation programs have been completely abandoned or are seldom used. Examples include: AGS, ALIGN, COMFORT, DESIGN, DIMAD, HARMON, LEGO, PETROS, RACETRACK, SYNCH, TRACY, TRANSPORT, TURTLE, UAL to name a few\cite{old_code, chao2013handbook}.

Oftentimes there is a huge impediment to maintaining these programs due to poorly-written code and lack of documentation. Additionally, many of the programs that are available tend to be ``rigid". That is, it is generally difficult to modify theses program to simulate something the program is not designed to simulate {\em a priori}. Adding a new type of lattice element that a particle can be tracked through is one such example.

Abandoned simulation programs represent a huge cost \cite{goble}, not only in terms of time and money spent in developing a program, but also in terms of researchers leveraging existing technology.  Indeed, a researcher who wants to simulate something that existing programs are unable to, will, due to time and monetary constraints, generally not be able to fully develop a comprehensive simulation program from scratch as compared to what could have been done if existing software could be leveraged.

A related problem involves programmers only considering the problem at hand during software development with the possible sharing of simulation data between programs only considered, if at all, as an afterthought. The result is that the ability to benchmark software, crosscheck results, perform regression tests, and do quality assurance testing for consistency is hindered. These types of checks are crucial in validating software accuracy and persistence.

\section{Modeling Needs}

As simulation programs become more complex due to the ever-increasing demands placed upon machine performance, the situation will become worse if not addressed. Such demands include the accelerator and beam physics Grand Challenges that have been identified recently by the community~\cite{LOI_ABPRoadmap,LOI_eva}
\begin{itemize}
\item  Increasing beam intensities by orders of magnitude.
\item  Increasing the beam phase-space density by orders of
magnitude, towards the quantum degeneracy limit.
\item Complete and highly accurate start-to-end “virtual particle accelerators” simulations.
\item Fast and accurate multi-objective optimization methods to speed up the design process.
\end{itemize}

Accelerator and beam modeling software development should allow for extensive testing of new functionality while preserving demonstrated capabilities on previously validated scenarios\cite{LOI_industry}. Performance and interoperability must be constantly improved to increase understanding (multi-physics problems) and optimization (machine learning). One must also ensure a transfer of knowledge over generations of scientists in form of formal education and easy accessibility to the tools.

Addressing these Grand Challenges will require a community effort to coordinate and modernize the current set of modeling tools, with capabilities that extend far beyond what the current toolset can do, including interdisciplinary simulations and advanced models for virtual prototyping of complete accelerator systems.

\subsection{Interdisciplinary Simulations}

Interdisciplinary simulations are important in a number of areas. In vacuo particle tracking coupled with particle/matter interactions is an example of a growing need. One application is in simulating the radiation induced by "dark current" electrons in accelerating cavities. This radiation may cause damage to cavities which leads to shortened lifetimes of the devices and a radiation safety hazard for the surrounding environment. Dark current induced problems have been observed at many facilities such as the CEBAF\cite{cebaf-dark}, LCLS-II\cite{lcls-dark}, ANL, etc. Sufficient shielding is required to properly contain the radiation which in turn requires a good understanding and prediction of radiation levels through simulations. Another example is the modeling of positron production in a target from the impact of high-energy electron beams accelerated through a linac injector. These simulations require accurate calculations using electromagnetic RF codes for accelerator structures and beam dynamics codes for particle transport in a beamline to characterize the beam profile before it hits the accelerator enclosure or the target. 

Particle/matter simulation codes exist. Examples include Geant4\cite{geant4}, FLUKA\cite{Ferrari:2005zk}, and MARS\cite{Mokhov:2017klc} which have traditionally been used for detector simulation in HEP experiments. However, since these codes and accelerator simulation codes have all been developed without common standards, interfacing them is a laborious task. A seamless simulation requires the proper transfer of field and particle data from accelerator to radiation codes. Communication in a standardized format such as openPMD\cite{openPMD}, which has been adopted in some accelerator codes, would help ensure efficient and error-free field and particle data transfers\cite{LOI_standards}. Another issue with an integrated simulation is in matching of the geometry of the vacuum chamber surface. The surface geometry in accelerator simulations is generally poorly defined if at all. The most comprehensive simulations define the surface using a finite element mesh generated from a CAD model. In contrast, radiation codes generally employ a faceted representation of the CAD model boundary. A converter for mapping finite element curved surfaces to faceted divisions on an interface boundary is required to accurately determine the location of a particle crossing from one computational domain to another. Much time and effort would be saved if the surface geometry descriptions were standardized so that a single converter module could be used in multiple codes.

Increasingly, accelerator simulation tools are also incorporating more micro-physics models to better describe the complex interplay of the various physics phenomena. One particular example being the emission modeling of a high brightness electron photocathode gun. A photocathode gun provides a high brightness electron source for the downstream accelerator beamline where the beam brightness can only be degraded, not improved. Thus, it is essential to understand the cathode emission characteristics and the method to control the beam quality in the gun environment through validated simulations. While Monte-Carlo photo electron emission simulations have been widely employed in studying photocathode performance for dedicated experiments, its potential in integrated simulations has only been explored recently~\cite{LOI_physics_based_injector_modeling}. For such purposes, a tight integration of the micro-physics models into existing gun simulations can be achieved via the best practices and standardization as discussed below.


\subsection{Realistic Models for Virtual Prototyping of Complete Accelerator Systems}

The increasing demand that accelerator models faithfully predict the performance of future facilities requires the development of more realistic simulation models. Here, two examples are given.

The first example is from single-particle nonlinear dynamics: Many accelerator codes use idealized models of beamline elements such as quadrupoles, sextupoles, bends, etc. The simplest models omit fringe fields. Better models are based on a fringe field that is a step function longitudinally. Unfortunately these models contain some terms that are infinite in the hard-edge limit, and codes with hard-edge models typically set these terms to zero\cite{ryne-map}. A better approximation is to assume some smooth analytical form for the fringe field. Though this approach is an improvement over the simpler models, it is still an idealization and there is no reason to expect that all of its nonlinear properties will precisely match those of the physical beamline element.

The precise prediction of single-particle nonlinear dynamics in accelerators can be accomplished using surface methods \cite{LOI_surfacemethods}. These methods have been known for many years but are not yet in widespread use in the accelerator community. The main idea of a surface method is to measure or numerically model the fields of a beamline element on a surface near but within the beam pipe. From there, the fields can be extrapolated inward and are represented by so-called generalized gradients that satisfy Maxwell’s equations. In the process, measured or computed errors in the fields at the surface are damped, leading to an accurate representation of the generalized gradients in the beam region. The generalized gradients can then be used to compute realistic transfer maps. These methods are now starting to gain popularity \cite{ipac2021_borland,PhysRevAccelBeams.23.104002,manikonda2006,manikonda2005}.

The second example is from the modeling of collective effects: The accurate simulation of 3D radiative phenomena, including coherent synchrotron radiation (CSR). Poisson solvers are essential to modeling space-charge effects in high intensity beams. In fact, there has been tremendous progress in modeling space-charge effects, and 3D parallel Poisson solvers are now ubiquitous. But despite advances in space-charge modeling, the understanding and modeling of 3D radiative phenomena remains an open problem. Most simulations involving CSR use a 1D model\cite{SALDIN1998}. More complicated 2D, 2.5D, and even some 3D models exist. However, these models, particularly the 3D models, sometimes involve questionable approximations and are extremely slow. There is presently no code that can accurately and reliably model 3D CSR effects including transient effects with a performance that makes it useful as a beamline design tool. As accelerators push boundaries with higher brightness beams, the ability to accurately model 3D CSR will become a key issue. This is true for accelerators with very high peak current beams such as those in beam-driven and laser-driven plasma accelerators. It is also true for future X-Ray free electron lasers (FELs) where complicated beam gymnastics involving bright electron beams is used to prepare the beam prior to entering the FEL.

The preceding describes just two examples among many of the need for more realistic models of accelerator components and phenomena. The development and application of new, more realistic models would enable virtual prototyping of entire accelerator systems including their nonlinear properties. It would allow the precise prediction of important properties -- nonlinear dynamics,  the thresholds for collective instabilities, etc. -- before a beamline is constructed and reduce the need for magnet shimming, the use of nonlinear correctors, etc. Ultimately such advanced capabilities would reduce cost, reduce risk, and improve the performance of future accelerator facilities. 

\section{Software Sustainability}

\begin{figure}[htb]
    \centering
    \includegraphics[width=5in]{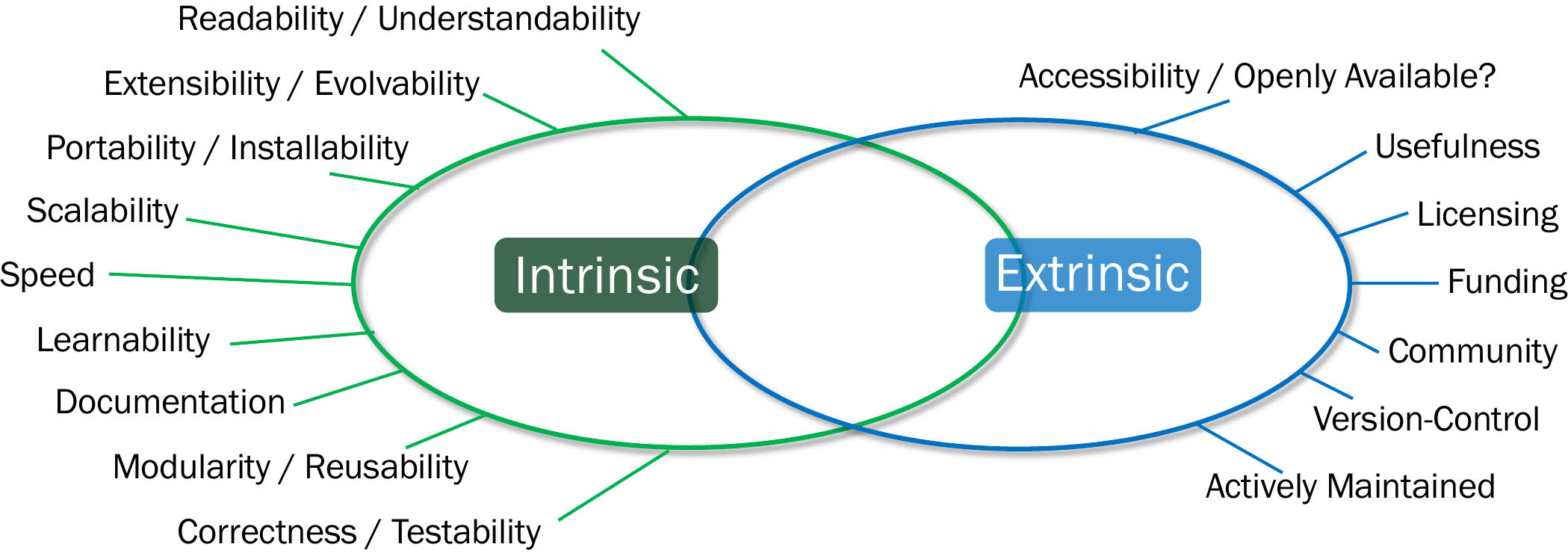}
    \caption{There are a number of aspects that make software sustainable. Broadly, they can be grouped into the ``intrinsic'' characteristics of the software itself and the ``extrinsic'' environment in which the software is developed and used.}
    \label{f:ss}
\end{figure}

There are several aspects that must be addressed to enable the development of the quality software that will be needed for the machines of tomorrow. One facet can be put under the rubric of ``software sustainability'' which can be defined as~\cite{katz}:
\begin{quotation}
``the capacity of the software to endure. In other words, sustainability means that the software will continue to be available in the future, on new platforms, meeting new needs.”
\end{quotation}

There are many aspects to software sustainability, as illustrated in Fig~\ref{f:ss}. Broadly, these aspects can be grouped into the ``intrinsic'' characteristics of the software itself and the ``extrinsic'' environment in which the software is developed and used. Software sustainability has been studied academically and there is even a Software Sustainability Institute~\cite{ssi}, which promotes ``the advancement of software in research by cultivating better, more sustainable, research software to enable world-class research''.

As mentioned above, many software packages developed for simulating accelerators,
while showing excellent results for their specific application and their era,
are not ``sustainable'' in the long run. However, software sustainability is extremely important given the limited resources 
that the accelerator community has for code development in conjunction with the even more limited resources 
for maintaining codes. To meet future needs, it is imperative that there is a community wide effort to promote sustainable practices.

\subsection{Software Toolkits}\label{sec:toolkits}

One aspect of developing sustainable software is the creation of software ``toolkits".
A software toolkit is an integrated set of modular routines that are used to develop and maintain applications.
As well as strengthening interoperability and sustainability of simulation software, a toolkit makes it possible to develop new programs in less time, with less effort, and with fewer bugs. 
There are many advantages to organizing software via a toolkit. This includes:
\begin{itemize}
\item 
Increased safety, since modular code provides a firewall. For example, a buggy module introduced into the toolkit will not affect programs that do not use it.
\item
A greater chance that bugs will be spotted since code modules get reused in different programs and therefore get greater scrutiny.
 \item
 By having modules that can read and write lattice information and data, the sharing of information between programs is made easier.
 \end{itemize}

A well-known example is Geant4~\cite{geant4}, which is a toolkit for the simulation of the passage of particles through matter. Geant4 has helped many researchers solve problems that would not be possible if a given researcher had tried to write the simulation code from scratch.

It is envisaged that toolkits would be developed for different accelerator physics purposes~\cite{LOI_toolkit}. As described in more detail in other sections, an important facet to maximizing sustainability would be to have accelerator community-wide policies and standards~\cite{LOI_standards} that would then enable the meshing of the toolkits into an ecosystem~\cite{LOI_ecosystem}, making it easier to develop the start-to-end simulations that are needed for the next generation of machines~\cite{LOI_eva}.

Along with toolkit development, there would be a need to develop general purpose, extensible simulation programs that can do the common tasks that accelerator physicists routinely do such as Twiss and orbit calculations, nonlinear optimization, lattice design, etc. That is, a program roughly equivalent to what programs like Tao\cite{tao}, MAD\cite{MAD2005}, or Elegant\cite{elegant} alleviates the need for a researcher to have to do programming when the researcher only wants to do a common task like calculating the closed orbit.

Besides the aforementioned general accelerator physics components, it is also important to have components for specific phenomena. Consider, for example, a Poisson solver library for modeling space-charge effects in high intensity/high brightness beams~
\cite{LOI_Poisson}. Subject to different boundary conditions, those Poisson solvers involve different numerical methods
that solve Poisson’s Equation on a grid. For an open boundary condition, a FFT-based Green’s function method could be used. For a closed boundary condition with regular shape, a finite difference spectral method could be used. For a closed boundary condition with irregular shape, a multigrid finite difference method is often used.
Rather than re-implementing the same functionalities (FFT-based or multigrid Poisson solver) multiple times in separate, incompatible and uncoordinated codes, the community would benefit greatly from consolidation in the development of a few selected toolkits for reuse across codes.

Accelerator simulation toolkits already exist: Accelerator Toolbox \cite{at}, Bmad \cite{bmad}, Cosy Infinity \cite{cosy}, Merlin++ \cite{merlin}, Warp \cite{warp}, IMPACT\cite{QIANG2000434} and FPP/PTC \cite{ptc}. Moving forward, one option would be to further develop existing toolkits. Another approach that would strengthen long-term sustainability would be to develop a new toolbox based on modern software practices. To not ``reinvent the wheel", this new toolbox should reuse existing algorithms and code wherever appropriate~\cite{LOI_ecosystem}.

\subsection{Software Ecosystem --- Interoperability and Policies}

Simulations of accelerator facilities can extend far beyond the accelerator itself. For example, in an XFEL, a comprehensive simulation will go from the creation of the X-ray pulses produced by electrons in an undulator, all the way to the simulation of the X-ray experiment. This includes X-ray transport, photon-sample interaction, signal transport, detector response, and data analysis. Realistically, no single software package or toolkit would cover all of these domains.

A software ``ecosystem'' is the set of libraries (toolkits) and applications that are developed somewhat independently but with a common set of rules and standards~\cite{LOI_ecosystem,LOI_standards}. An ecosystem facilitates the implementing of frameworks for start-to-end workflows that span over multiple components. Recently developed examples of such frameworks include LUME~\cite{lume}, PaNOSC~\cite{panosc}, and SIMEX~\cite{simex}. Such integration platforms aid in the rapid adoption for new design and analysis needs, using complex tools in automated workflows such as needed for AI/ML and optimization.

As of today, data exchange between individual applications is often based on widely supported file formats that implement a common meta-data schema, see section~\ref{sec:standards}.
One of the opportunities with a compatible software ecosystem would be the further adoption of such standards as well as abstraction of low-level data layouts, which would foster innovative combinations of toolkits with bespoke software into new applications for tightly coupled models.
A large community of developers can focus on domain-specific needs and help improve software capability and sustainability across projects, while reducing maintenance and development time due to common practices and fundamentals developed by specialists.
Fundamental building blocks would also address performance and portability needs in low-level libraries. This could include support, for example, of, CPUs, GPUs, multi-node communication patterns, and multi-platform compiling.
Inefficiencies and bugs are spotted earlier due to re-use.
A progressive path for both adoption and continued modernization of such components is possible, starting from existing libraries.

Beyond the data exchange through common file formats and meta-data schemata, it is sometimes desirable to exchange data among the simulation tools in a more integrated, finer-grained manner that may be at the level of each simulation time step\cite{1591190,Nagaitsev2021,LOI_physics_based_injector_modeling}. This may happen when multiple physics effects are to be simultaneously simulated or when coupling multiple solvers, where data exchange can be either in single direction or bi-directional. Such an in-memory or cross-node data transfer can be made to respect physics constraints (conservation, divergence free conditions, etc.) and is also much more efficient than file I/O based transfers\cite{Zhang2017InmemorySA,7877157,alpine2017}. This need can be best served with the common low-level data layouts and interfaces as mentioned above and/or by promoting/enabling interoperable mesh/particle capabilities among existing tools in the ecosystem.   

In order to advance accelerator modeling further in the direction of a compatible, extensible ecosystem, some general standardization needs, including best practices in software development, have been identified~\cite{LOI_ecosystem}.
These are based on community policies that have been established over years by teams of specialists in scientific software development and computing.
For instance, the accelerator community could be leveraging on the Interoperable Design of Extreme-scale Application Software (IDEAS) project \cite{IDEAS} and its Extreme-scale Scientific Software Development Kit (xSDK) \cite{xSDK}.

An ecosystem would be best developed on the basis of permissive open source licenses to allow reuse, cross-institutional and international collaboration and adoption.
For vertical software integrations, wide-ranging open source libraries from numerical solvers to optimizers (e.g., Hypre~\cite{HYPRE}, libEnsemble~\cite{libEnsemble}), various meshing and mesh-refinement (AMReX~\cite{AMReX}), and particle (for example, CoPA Cabana \cite{CopaCabana}) and mesh/particle remapping (e.g., Portage~\cite{Portage}) libraries could be combined as a foundation for the accelerator and beam physics components.

\subsection{I/O Standardization}\label{sec:standards}

Traditionally, existing accelerator modeling applications were driven by individual groups with little coordination between development activities. This is detrimental for a number of reasons including hindering the ability to benchmark the software, and crosscheck results.
Indeed, it is the consensus within the computer science and majority of computational science sub-domains that computational results need to be reproducible and independently replicable\,\cite{REANA}.

As accelerator software becomes more complex, and there is a strong desire to integrate capabilities (such as new methods, start-to-end modeling, common analysis needs and machine-guided optimization), standards for data and workflows can help increase productivity and sustainability \cite{LOI_standards}.
Recently, activities such as those supported by the Consortium for Advanced Modeling of Particle Accelerators (CAMPA)~\cite{CAMPA}, helped coordinate the standardization of data exchange and simulation control, with the aim of connecting large existing frameworks and enabling innovative workflows.
Such standardization efforts can also provide the basis for an integration into scientific data portals to curate and re-use modeling data\,\cite{CERNdata,FAIRdata,ATL-PHYS-PUB-2019-032}. Besides improving reproducibility of simulations and preserving value of collected data, such efforts can also aid meta-studies and tests for new theories.

With respect to accelerator simulation data, historically, there exists a variety of low-level data file formats, which individual modeling tools use with custom-made meta-data conventions to express the domain-specific data consumed and generated by individual tools.
Yet new, efficient data formats and highly-tuned I/O libraries are continuously being developed by computer scientists and existing paradigms, such as POSIX file-I/O, are overtaken by modern approaches such as data streams, object storage and relaxed I/O-constrains in highly-parallel computing environments.
Adoption of these low-level file formats for accelerator modeling is needed to utilize the progress in modern storage and data transport technology and overcome bottlenecks arising from file-based storage of high-fidelity data as well as manual data analysis and curation efforts.

The Open Standard for Particle-Mesh Data Files (openPMD)\,\cite{openPMD} successfully demonstrates that defining compatible meta-data in a file-format agnostic organization for accelerator and beam data is possible, while using scalable, modern file formats from computer science \cite{ADIOS2,HDF5,Huebl2017}.
OpenPMD is organized around a written, versioned text document that is supported by tooling for validation, examples, a project catalogue, libraries and programs, which all have their respective documentation and tutorials.
Individual compatible projects, software, and data are published by a variety of authors\,\cite{openPMDprojects}.

Complementary to openPMD, the Standard Input Format for Particle-In-Cell Codes (PICMI) addresses the challenge of unified simulation design by defining a common input layer that focuses on the physical description of a problem set\,\cite{PICMI}.
Currently, PICMI is implemented as a high-level Application Programming Interface (API), which is an approach similar to successful community math and HPC APIs, for example, BLAS for linear algebra and MPI for multi-node message passing on supercomputers.
The programming language used for the PICMI API is Python, which is a well-known scripting language suitable for rapid simulation design and backed by a vast, extensible software ecosystem.

\section{Advanced Concepts and Future Computing}

\subsection{Advanced Concepts}

Advanced Accelerator Concepts (AAC) offer accelerating gradients that go beyond the limitations of standard RF technologies, sometimes by an order of magnitude or more, leading to the prospect of much more compact - and in some cases proportionally cheaper - technologies. AAC includes laser-driven plasma acceleration (aka LWFA or LPA), charged-particle-beam-driven plasma acceleration (aka PWFA), structure-based wakefield accelerators (SWFA), dielectric laser acceleration (DLA) or laser-ion acceleration. While the modeling of DLA involves mostly simulation tools that are already used for conventional accelerators, the modeling of the other schemes involve different models (and thus simulation tools) and can be significantly more challenging computationally, in particular for those involving plasmas. The description of the specific needs and challenges of these tools is given in another paper of this issue \cite{VayICFA2021}. Aside from the different physics, numerical methods and computational needs, other challenges and solutions are essentially the same as for conventional accelerators with regard to, e.g., software sustainability, standardization, validation, verification, usability, integrated workflows and frameworks, toolkits, ecosystems, centers, evolving and future architectures (including quantum computing) that are described in this paper.

In terms of numerical approaches, advanced concepts include the need to investigate novel numerical methods and approaches beyond Particle-In-Cell (PIC)  methods~\cite{Friedman:1992,Hockney1966ComputerSU,QIANG2000434,AMUNDSON2006229} that have shown promise for future incorporation into simulation software as modules.  These approaches could potentially tackle different types of advanced simulation problems.  Examples include symplectic multiparticle
space-charge simulations~\cite{PhysRevAccelBeams.20.014203,PhysRevAccelBeams.21.054201}, Fast Multipole Method (FMM)-based approaches~\cite{GREENGARD1987325,10.1145/1048935.1050165,10.1145/2160718.2160740,ZHANG2011338,LGZ1-2011}, boundary integral solvers~\cite{BIROS2004317,KLOCKNER2013332,MORSE2021110511} and hybrid solvers such as {\em Vico-Greengard-Ferrando}~\cite{10.1016/j.jcp.2016.07.028,zou2021fftbased}.

\subsection{Evolving architectures in standard computing}

Developments in existing and planned DOE High Performance Computing (HPC) facilities indicate an increasing reliance on architectures with attached co-processors, styled as GPU computing.
These processors offer an attractive boost in computing performance with respect to any of the common denominators such as hardware cost, electricity cost, or wall clock time to accomplish a particular workload.
This boost in performance is achieved by implementing highly parallel computational engines acting on localized data at the expense of general purpose computing capabilities.
Computational problems whose algorithms can be cast in this paradigm benefit greatly from this style of computing architecture.

Fortunately, many accelerator simulation problems are within this class.
Propagating many independent particles through beamline components is seemingly custom made for this kind of computing.
The issue facing the field then is that programming these devices requires specialized techniques.
Typically, data for computations should be transferred to co-processor memory and arranged carefully for efficient parallel processing.
This transfer is usually slow and should be minimized or eliminated over the course of a long computation.
It may not be possible to implement some current algorithms on co-processor hardware; new algorithms will need to be developed.

Computing on current general purpose processors will continue to be an important part of the landscape.
Software should be built to run on either standard hardware or new co-processor based architectures, of which there are several.
Although the most well-established solutions for GPU computing are provided by NVidia, AMD and Intel are also building HPC systems.
Some accelerator simulation codes have been implemented in CUDA, the NVidia specific GPU programming language which is specific to NVidia GPUs.
Ideally, the accelerator simulation community would avoid dependence on a single co-processor supplier.
Fortunately, there are several major software efforts to build platform-independent parallel computing frameworks that can support execution on either CPU or GPU processors with a single high-level code base.
No particular framework is clearly superior to the others, and each is targeted for use in a particular language (C++ or Fortran) and level of abstraction.
The important lesson is that the community and maintainers of import simulation tools should be supported in either upgrading or re-implementing their algorithms for use in the near and medium future on what will be the dominant scientific computing architecture.

\subsection{Adaptively Tuned Simulations as Online Virtual Diagnostics}

Some of the most detailed accelerator diagnostics are X-band transverse deflecting cavities. However, deflecting cavities are limited to a resolution of 1 $\mu$m or 3.3 fs/pixel, and destroy the beam during the measurement process \cite{behrens2014few}. This measurement floor is of concern since beams at advanced plasma wakefield accelerators and free electron laser (FEL) facilities are starting to exceed those limits. The Facility for Advanced Accelerator Experimental Tests (FACET-II) will provide bunch lengths as low as (1 $\mu$m or $\sim$3 fs) at 12 GeV \cite{joshi2018plasma}, attosecond two-color X-ray pulses have been achieved at the SwissFEL \cite{malyzhenkov2020single}, and designs for the international linear collider call for 3.2 nC bunches with 30 $\mu$m bunch lengths at 250 GeV \cite{aihara2019international}.

As bunch lengths decrease beyond the resolution of existing diagnostics, simulations will be relied on as virtual diagnostics. Simulation-driven approaches are common in other scientific fields, such as coherent diffraction imaging where physics simulations translate 2D X-ray diffraction intensity measurements to 3D electron densities of crystals. Using simulations as online diagnostics requires a close match with accelerator performance. This is a non-trivial problem even for particle-tracking codes with millions of macro-particles because once accelerators are built they do not perfectly match the designs that simulation models are based on. During installation misalignments are introduced, the electromagnetic fields of components do not perfectly match simulated fields, and once operational the components and the initial input beam distributions drift with time and are perturbed by disturbances not accounted for.

Closely matching simulations to accelerators requires adaptive feedback. Diagnostics can be compared to simulation-based predictions and simulation parameters can be tuned in real-time to achieve a match between measurements and simulation outputs. Once a match is achieved, because of physics constraints within the simulation, it is likely that other beam properties are uniquely matched. Recently a LiTrack model of FACET was adaptively tuned online to match the simulated beam's energy spread spectrum $\hat{\rho}_E(x,t)$ to its measured $\rho_E(x,t)$, minimizing the error $e(t) = \int \left | \hat{\rho}_E(x,t) - \rho_E(x,t) \right | dx$, to track the time-varying longitudinal phase space $(z,E)$ of the electron beam \cite{bane2005litrack,scheinker2015adaptive}. Efforts are also underway to utilize ML tools to map diagnostics back to input beam distributions to be used as the initial conditions of accelerator models \cite{scheinker2021adaptive}. Such approaches are possible with any simulation tool for tracking time-varying beams. By taking advantage of GPUs and field programmable gate arrays it may be possible to use such adaptive models as virtual diagnostics in real-time shot-to-shot for high repetition rate machines.

\subsection{Getting ready for the arrival of quantum computing age}

Studies of collective effects are essential for modern accelerators with high intensity beams. The start-to-end simulation of an accelerator using real beams with billions or more particles remains challenging and expensive even with state-of-the-art exascale machines. The development of quantum computers provides new opportunities to enhance the particle accelerator community's simulation abilities. 

A quantum computer is a device that utilizes the special properties of quantum mechanics to perform computation, which can potentially provide an exponential improvement in efficiency for some classes of simulations. The three properties of quantum superposition, interference, and entanglement of a quantum state make quantum computing different from classical computing. Due to quantum superposition, the information stored in a quantum system scales exponentially with the number of qubits as opposed to the linear scaling with the number of bits in a classical system. Both commercial companies and research institutes are developing techniques for building quantum computers. Current state-of-the-art quantum computers have above 50 qubits and quantum supremacy has been demonstrated on some specific problems \cite{arute2019quantum,zhong2020quantum}.  Quantum computing is currently available to the public through cloud services provided by some commercial companies such as IBM \cite{IBMQexperience}, D-Wave \cite{DWaveLeap2}, Amazon \cite{AmazonQSulution} and Microsoft \cite{AzureQ}. Meanwhile, in academia, studies on quantum algorithms have also been booming in the past few years. The work that is most relevant for accelerator modeling relates to solving linear systems \cite{harrow2009quantum,clader2013preconditioned,childs2017quantum,bravo2019variational,lee2019hybrid} as well as ordinary differential equations (ODEs) and partial differential equations (PDEs) \cite{leyton2008quantum,berry2014high,arrazola2019quantum,childs2020high,xin2020quantum}. Solving Poisson’s equation and the Vlasov equations -- which are often used in the simulation of collective effects such as space charge, the beam-beam interaction, and coherent synchrotron radiation -- with quantum computing is being explored \cite{cao2013quantum,wang2019quantum, engel2019quantum}. Quantum computer simulators, as code developing-, debugging-, and testing platforms, are available for almost all mainstream programming languages, e.g. C/C++, Python, Java, Matlab, etc. There also exist some languages specifically designed for quantum computing \cite{MicrosoftQDev}. All these provide the community with the fundamental blocks to build simulation tools for beam and accelerator physics. 

Most previous work on quantum algorithms focused on the realization of an algorithm with quantum circuits, but applying the algorithm to solve a practical scientific problem is seldom discussed \cite{dodin2020applications}. Clearly there is a gap between the development of the algorithms and their implementation. To make a problem suitable for quantum simulations, it has to be described by unitary operators so that the system, which may be classical, is mathematically equivalent to a quantum system. Additionally, in quantum computing, all variables are stored in quantum states. Preparing the initial states and reading out the results accurately can be time intensive. The traditional way of using a large number of particles in simulation and producing all of their coordinates as output  would be unpractical. 

Given the above considerations, at least in the near future, a quantum computer will not replace a classical computer but will probably work together with one. It is thus probably better for now to focus on how to make new quantum simulation tools that collaborate with the existing pool of accelerator modeling programs. 
Ideally a protocol will be invented through which the quantum packages could be called by the classical programs. 
Also needed are innovative analyzing tools to process the simulation results, which may happen before the reading-out.
To achieve these tasks, the accelerator physics problems that will benefit from quantum computing need to be identified and for each of them the proper mathematical model will need to be established, which may be different from the conventional classical one. For this, contributions from experts in both accelerator physics and quantum computing are required.  

\section{Centers for Accelerator and Beam Physics Modeling}

It quickly becomes clear that, in order to achieve what is described and 
proposed in the other sections, a coherent and consolidated
effort is needed. This is best achieved in the form of dedicated Centers 
for Accelerator and Beam Physics Modeling \cite{LOI_center}.
Other areas of computer science have already
embraced this fact. New colleges for computing are established at universities
to consolidate the dispersed computing efforts of the various departments 
(e.g. MIT's Schwarzman College of Computing \cite{schwarzman}), 
and new centers for Quantum Computing \cite{quics, ibm:quantum} 
have been built. Exascale computing has been embraced through the Exascale 
Computing Project \cite{exascale}. 
The US Department of Energy (DOE) has founded SciDAC \cite{scidac} to accelerate progress in
scientific computing across the different programs supported by DOE: Advanced Scientific Computing Research, Basic Energy Sciences, Biological and Environmental Research, Fusion Energy Sciences, High-Energy Physics, and Nuclear Physics.

The respective communities have benefited strongly from these new centers and the partnerships across disciplines.

Accelerator and Beam Physics Modeling would no doubt benefit similarly. The 
centers can be at a given location or distributed geographically and among 
institutions across laboratories, academia and industrial partners. 
They would bring together domain scientists (computational accelerator and 
beam physicists), applied mathematicians, computer scientists and software engineers 
with collaborations across the full landscape of accelerator modeling. 
In addition, some of the computer science centers mentioned above are already 
supporting accelerator modeling efforts on which the Centers for Accelerator and Beam Physics Modeling could build.

Depending on the overall size, the centers could enable part or all of the following:
\begin{itemize}
\setlength\itemsep{0.1em}
    \item Community development and maintenance of codes using industry-standard quality processes by dedicated, specialized teams \cite{LOI_industry}.
    \item Collect libraries for field solvers, particle trackers, and other modules.
    \item Provide a modular community ecosystem for multiphysics particle accelerator 
          modeling and design \cite{LOI_ecosystem}.
    \item Standardize input scripts, output data, lattice description and 
          start-to-end workflows \cite{LOI_standards}.
    \item Provide compatibility layers to use the same libraries and modules in a 
          number of programming languages.
    \item Development and maintenance of End-to-end Virtual Accelerators (EVA) \cite{LOI_eva}. 
    \item User support, high-quality and detailed documentations, 
          online tutorials, and training.
    \item Easy-to-use, standardized, user interfaces for preparation and analysis of simulations.
    \item Automated tools for ensemble simulations for optimization with builtin AI/ML support. 
    \item Suite of test problems with well-characterized solutions for benchmarking, quality assurance and regression testing.
    \item Development, analysis and efficient implementation of novel 
          algorithms and numerical methods (e.g., high-order solvers, symplectic multiparticle tracking, Fast Multipole Methods,
          adaptive mesh refinement).
    \item Providing a space to meet (physically or virtually) for the integration 
          of developments from contributors into larger codes, such as PhD projects 
          from external groups, organizing development hackathons, 
          knowledge-transfer, and onboarding.
    \item Developing and organizing workshops for developers and users of codes alike.
          Inviting national and international speakers/developers (travel/hosting funds).
    \item Interacting with existing schools, by developing and maintaining 
          state-of-the-art educational resources (e.g. tutorials, lectures) on codes.
    \item Exploration of novel use of machine learning for accelerator modeling, 
          and, further in the future, of quantum computing \cite{LOI_ML}.
\end{itemize}

Multiple Centers can be organized through a Consortium (e.g., CAMPA \cite{CAMPA}). Except for special restrictions such as export control, it would be desirable for the software developed by the Center to be open source, enabling crosschecking, testing and contribution by the community at large, beyond the participants to the Center(s) \cite{LOI_open}.

\section{Conclusion}

The historically disorganized nature of accelerator software development has been a major impediment to creating the quality applications that are needed to both run existing machines as well as to design future ones: 

\blockquote{{\it Computer simulations play an indispensable role in all accelerator areas.
Currently, there are many simulation programs used for accelerator physics.
There is, however, very little coordination and cooperation among the developers of these codes.
Moreover there is very little effort currently being made to make these codes generally available to the accelerator community ...}

\hfill --- \textup{HEPAP report}, 2015}

Consider the case of the Superconducting Super Collider (SSC)\cite{tunnelvisions}. The SSC was terminated, after millions of dollars were spent, in part due to a flawed simulation. The decision to increase the vacuum chamber bore diameter from 4~cm to 5~cm was based in part upon a faulty simulation using the code SSCTRK\cite{ritson1990}. This code was newly developed for the SSC\cite{ssctrk} and had not been thoroughly vetted at the time the decision was made. After refinements to the Monte Carlo error model, and after upgrading the SSCTRK code, it was realized that the original 4~cm bore had been adequate. The unnecessary enlargement to 5~cm lead to cost increases along with some turmoil caused by having to redesign vacuum chambers, magnets and other machine components. Ultimately, this was was a major factor in the demise of the SSC.

Communities in the computational sciences face similar challenges, and physics modeling groups would benefit from incorporating research results, best practices, participate actively in aforementioned bodies, and anticipate trends in the broader computer science and computational science community. In sustainable workflows, one wants to avoid heroic efforts relying on few individuals, and instead provide an environment that is inclusive and thrives with contributions from various educational backgrounds.
Physics groups that integrate early with external computer science, applied mathematics and computational physics efforts can share modular solutions\,\cite{LOI_ecosystem}, drive cross-domain visions and avoid missing out on solutions developed in related scientific domains.

There have been notable improvements in recent years. Some are centered around novel, open source particle-in-cell codes for laser-plasma modeling and around collaborations on standards\,\cite{LOI_standards}.
Yet other projects are essentially walled gardens with varying access levels to simulation programs, their source code, documentation, support, and usage rights.
Another challenge lies in the publication of computational work. Analysis routines, source codes, inputs, and simulation data are often not 
openly archived in sufficient detail, which hinders reproducibility and adoption of published methods by other groups.
This situation poses a significant risk and calls for an advancement of modeling practices that can adequately address the needs of decade-long basic science projects.

One part of the problem is that, traditionally, performance metrics of scientific success aim solely on publication numbers. The extra time and effort to make code sustainable is, by this criterion, unproductive. The end result, however, is extensive waste. It is imperative that funding for software sustainability be made available to the accelerator community. Ultimately, such funding will pay for itself many times over. 

\bibliographystyle{unsrtnat}
\bibliography{biblio}

\begin{thebibliography}{116}
\providecommand{\natexlab}[1]{#1}
\providecommand{\url}[1]{\texttt{#1}}
\expandafter\ifx\csname urlstyle\endcsname\relax
  \providecommand{\doi}[1]{doi: #1}\else
  \providecommand{\doi}{doi: \begingroup \urlstyle{rm}\Url}\fi

\bibitem[old()]{old_code}
Accelerator physics codes.
\newblock URL \url{https://en.wikipedia.org/wiki/Accelerator_physics_codes}.

\bibitem[Chao et~al.(2013)Chao, Mess, et~al.]{chao2013handbook}
Alexander~Wu Chao, Karl~Hubert Mess, et~al.
\newblock \emph{Handbook of accelerator physics and engineering}.
\newblock World scientific, 2013.

\bibitem[Goble(2014)]{goble}
Carole Goble.
\newblock Better software, better research.
\newblock \emph{IEEE Internet Computing}, 18, 2014.

\bibitem[Nagaitsev et~al.(2020)Nagaitsev, Huang, J., Vay, Piot, Spentzouris,
  Rosenzweig, Cai, Cousineau, Conde, Hogan, Valishev, Minty, Zolkin, Huang,
  Shiltsev, Seeman, Byrd, and Patterson]{LOI_ABPRoadmap}
S.~Nagaitsev, Z.~Huang, Power J., J.-L. Vay, P.~Piot, L.~Spentzouris,
  J.~Rosenzweig, Y~Cai, S.~Cousineau, M.~Conde, M.~Hogan, A.~Valishev,
  M.~Minty, T.~Zolkin, X.~Huang, V.~Shiltsev, J.~Seeman, J.~Byrd, and J.R.
  Patterson.
\newblock {Accelerator and Beam Physics: Grand Challenges and Research
  Opportunities}.
\newblock \emph{Snowmass21 LOI}, 2020.
\newblock URL
  \url{https://www.snowmass21.org/docs/files/summaries/AF/SNOWMASS21-AF1_AF7_S_Nagaitsev-056.pdf}.

\bibitem[Vay et~al.(2020{\natexlab{a}})Vay, Sagan, Huebl, Th\'evenet, Lehe, Ng,
  Vincenti, Bussmann, Debus, Pausch, and Qiang]{LOI_eva}
Jean-Luc Vay, David Sagan, Axel Huebl, Maxence Th\'evenet, Rémi Lehe, Cho-Kuen
  Ng, Henri Vincenti, Michael Bussmann, Alexander Debus, Richard Pausch, and
  Ji~Qiang.
\newblock {End-to-End Virtual Accelerators (EVA)}.
\newblock \emph{Snowmass21 LOI}, 2020{\natexlab{a}}.
\newblock URL
  \url{https://www.snowmass21.org/docs/files/summaries/CompF/SNOWMASS21-CompF2_CompF0-AF1_AF0_Vay-067.pdf}.

\bibitem[Lehe et~al.(2020{\natexlab{a}})Lehe, Huebl, Vay, Friedman, Thevenet,
  Mitchell, Bruhwiler, Grote, Cowan, Vincenti, Hanuka, Cros, Yoffe, Widera,
  Bussmann, and Edelen]{LOI_industry}
Remi Lehe, Axel Huebl, Jean-Luc Vay, Alexander Friedman, Maxence Thevenet, Chad
  Mitchell, David Bruhwiler, David Grote, Benjamin Cowan, Henri Vincenti, Adi
  Hanuka, Brigitte Cros, Samuel Yoffe, Rene Widera, Michael Bussmann, and
  Auralee~Linscott Edelen.
\newblock {Embracing modern software tools and user-friendly practices, when
  distributing scientific codes}.
\newblock \emph{Snowmass21 LOI}, 2020{\natexlab{a}}.
\newblock URL
  \url{https://www.snowmass21.org/docs/files/summaries/CompF/SNOWMASS21-CompF2_CompF0_Lehe-076.pdf}.

\bibitem[Hovater et~al.(2017)Hovater, Allison, Biallas, Bachimanchi, Daly,
  Drury, Freyberger, Geng, Lahti, Legg, Mounts, Nelson, Plawski, and
  Powers]{cebaf-dark}
C.~Hovater, T.~Allison, G.~Biallas, R.~Bachimanchi, E.~Daly, M.~Drury,
  A.~Freyberger, R.~L. Geng, G.~Lahti, R.~Legg, C.~Mounts, R.~Nelson,
  T.~Plawski, and T.~Powers.
\newblock Operation of the cebaf 100 mv cryomodules.
\newblock In \emph{Proc. of Linear Accelerator Conference (LINAC'16), East
  Lansing, MI, USA, 25-30 September 2016}, number~28 in Linear Accelerator
  Conference, pages 65--67, Geneva, Switzerland, 5 2017. JACoW.
\newblock ISBN 978-3-95450-169-4.
\newblock \doi{https://doi.org/10.18429/JACoW-LINAC2016-MOOP11}.
\newblock URL \url{http://jacow.org/linac2016/papers/moop11.pdf}.

\bibitem[Lewandowski et~al.(2015)Lewandowski, Field, Fisher, Nuhn, and
  Welch]{lcls-dark}
J.~R. Lewandowski, R.~C. Field, A.~S. Fisher, H.-D. Nuhn, and J.~J. Welch.
\newblock Rf gun dark current suppression with a transverse deflecting cavity
  at lcls.
\newblock In \emph{Proc. 37th Int. Free Electron Laser Conf. (FEL'15)}, pages
  583--586. JACoW Publishing, 8 2015.
\newblock \doi{doi:10.18429/JACoW-FEL2015-WEP001}.
\newblock https://doi.org/10.18429/JACoW-FEL2015-WEP001.

\bibitem[Allison et~al.(2016)Allison, Amako, Apostolakis, Arce, Asai, Aso,
  Bagli, Bagulya, Banerjee, Barrand, Beck, Bogdanov, Brandt, Brown, Burkhardt,
  Canal, Cano-Ott, Chauvie, Cho, Cirrone, Cooperman, Cortés-Giraldo, Cosmo,
  Cuttone, Depaola, Desorgher, Dong, Dotti, Elvira, Folger, Francis, Galoyan,
  Garnier, Gayer, Genser, Grichine, Guatelli, Guèye, Gumplinger, Howard,
  Hřivnáčová, Hwang, Incerti, Ivanchenko, Ivanchenko, Jones, Jun,
  Kaitaniemi, Karakatsanis, Karamitros, Kelsey, Kimura, Koi, Kurashige,
  Lechner, Lee, Longo, Maire, Mancusi, Mantero, Mendoza, Morgan, Murakami,
  Nikitina, Pandola, Paprocki, Perl, Petrović, Pia, Pokorski, Quesada, Raine,
  Reis, Ribon, {Ristić Fira}, Romano, Russo, Santin, Sasaki, Sawkey, Shin,
  Strakovsky, Taborda, Tanaka, Tomé, Toshito, Tran, Truscott, Urban, Uzhinsky,
  Verbeke, Verderi, Wendt, Wenzel, Wright, Wright, Yamashita, Yarba, and
  Yoshida]{geant4}
J.~Allison, K.~Amako, J.~Apostolakis, P.~Arce, M.~Asai, T.~Aso, E.~Bagli,
  A.~Bagulya, S.~Banerjee, G.~Barrand, B.R. Beck, A.G. Bogdanov, D.~Brandt,
  J.M.C. Brown, H.~Burkhardt, Ph. Canal, D.~Cano-Ott, S.~Chauvie, K.~Cho,
  G.A.P. Cirrone, G.~Cooperman, M.A. Cortés-Giraldo, G.~Cosmo, G.~Cuttone,
  G.~Depaola, L.~Desorgher, X.~Dong, A.~Dotti, V.D. Elvira, G.~Folger,
  Z.~Francis, A.~Galoyan, L.~Garnier, M.~Gayer, K.L. Genser, V.M. Grichine,
  S.~Guatelli, P.~Guèye, P.~Gumplinger, A.S. Howard, I.~Hřivnáčová,
  S.~Hwang, S.~Incerti, A.~Ivanchenko, V.N. Ivanchenko, F.W. Jones, S.Y. Jun,
  P.~Kaitaniemi, N.~Karakatsanis, M.~Karamitros, M.~Kelsey, A.~Kimura, T.~Koi,
  H.~Kurashige, A.~Lechner, S.B. Lee, F.~Longo, M.~Maire, D.~Mancusi,
  A.~Mantero, E.~Mendoza, B.~Morgan, K.~Murakami, T.~Nikitina, L.~Pandola,
  P.~Paprocki, J.~Perl, I.~Petrović, M.G. Pia, W.~Pokorski, J.M. Quesada,
  M.~Raine, M.A. Reis, A.~Ribon, A.~{Ristić Fira}, F.~Romano, G.~Russo,
  G.~Santin, T.~Sasaki, D.~Sawkey, J.I. Shin, I.I. Strakovsky, A.~Taborda,
  S.~Tanaka, B.~Tomé, T.~Toshito, H.N. Tran, P.R. Truscott, L.~Urban,
  V.~Uzhinsky, J.M. Verbeke, M.~Verderi, B.L. Wendt, H.~Wenzel, D.H. Wright,
  D.M. Wright, T.~Yamashita, J.~Yarba, and H.~Yoshida.
\newblock Recent developments in geant4.
\newblock \emph{Nuclear Instruments and Methods in Physics Research Section A:
  Accelerators, Spectrometers, Detectors and Associated Equipment},
  835:\penalty0 186 -- 225, 2016.

\bibitem[Ferrari et~al.(2005)Ferrari, Sala, Fasso, and Ranft]{Ferrari:2005zk}
Alfredo Ferrari, Paola~R. Sala, Alberto Fasso, and Johannes Ranft.
\newblock {FLUKA: A multi-particle transport code (Program version 2005)}.
\newblock 10 2005.
\newblock \doi{10.2172/877507}.

\bibitem[Mokhov and James(2017)]{Mokhov:2017klc}
Nikolai~V. Mokhov and Catherine~C. James.
\newblock {The MARS Code System User\textquoteright{}s Guide Version 15(2016)}.
\newblock 2 2017.
\newblock \doi{10.2172/1462233}.

\bibitem[Huebl et~al.(2015)Huebl, Lehe, Vay, Grote, Sbalzarini, Kuschel, Sagan,
  Pérez, Koller, and Bussmann]{openPMD}
Axel Huebl, Rémi Lehe, Jean-Luc Vay, David~P. Grote, Ivo Sbalzarini, Stephan
  Kuschel, David Sagan, Frédéric Pérez, Fabian Koller, and Michael Bussmann.
\newblock openpmd: A meta data standard for particle and mesh based data.
\newblock 2015.
\newblock \doi{10.5281/zenodo.591699}.
\newblock URL \url{https://doi.org/10.5281/zenodo.591699}.

\bibitem[Huebl et~al.(2020{\natexlab{a}})Huebl, Vay, Lehe, Th\'evenet, Mayes,
  Sagan, Tsai, E, Tsung, Vincenti, Pousa, Cook, Gessner, Poeschel, Bussmann,
  Grote, Murphy, Schmitz, Yoon, Bruhwiler, Cranmer, Yoffe, Cros, Edelen, and
  Stark]{LOI_standards}
A.~Huebl, J.-L. Vay, R.~Lehe, M.~Th\'evenet, C.~Mayes, D.~Sagan, Y.-D. Tsai,
  J.~C. E, F.~Tsung, H.~Vincenti, A.~Ferran Pousa, N.~M. Cook, S.~J. Gessner,
  F.~Poeschel, M.~Bussmann, D.~P. Grote, N.~A. Murphy, R.~Schmitz, C.~H. Yoon,
  D.~L. Bruhwiler, K.~Cranmer, S.~R. Yoffe, B.~Cros, A.~L. Edelen, and
  G.~Stark.
\newblock Develop/integrate data standards and start-to-end workflows.
\newblock \emph{Snowmass21 LOI}, 2020{\natexlab{a}}.
\newblock URL
  \url{https://www.snowmass21.org/docs/files/summaries/CompF/SNOWMASS21-CompF2_CompF7-AF1_AF0_Huebl-079.pdf}.

\bibitem[Huang et~al.(2020)Huang, Kwan, Pavlenko, Ng, and
  Wang]{LOI_physics_based_injector_modeling}
C.-K. Huang, T.J.T. Kwan, Vitaly Pavlenko, Cho-Kuen Ng, and Erdong Wang.
\newblock {Physics-based high-fidelity modeling of high brightness beam
  injectors}.
\newblock \emph{Snowmass21 LOI}, 2020.
\newblock URL
  \url{https://www.snowmass21.org/docs/files/summaries/AF/SNOWMASS21-AF7_AF1-CompF2_CompF0_Huang-183.pdf}.

\bibitem[Ryne and Dragt(1996)]{ryne-map}
R.~Ryne and A.~Dragt.
\newblock Numerical computation of transfer maps using lie algebraic methods.
\newblock In \emph{Proceedings of the 12th IEEE Particle Accelerator
  Conference.}, pages 1081--3, 1996.

\bibitem[Ryne et~al.(2020)Ryne, Abell, Berz, Bruhwiler, Dragt, Makino,
  Mitchell, Ng, Qiang, Vay, Venturini, and Walstrom]{LOI_surfacemethods}
Robert Ryne, Dan Abell, Martin Berz, David Bruhwiler, Alex Dragt, Kyoko Makino,
  Chad Mitchell, Cho-Kuen Ng, Ji~Qiang, Jean-Luc Vay, Marco Venturini, and
  Peter Walstrom.
\newblock {Surface methods for precision accelerator design and virtual
  prototyping of accelerator systems}.
\newblock \emph{Snowmass21 LOI}, 2020.
\newblock URL
  \url{https://www.snowmass21.org/docs/files/summaries/CompF/SNOWMASS21-CompF2_CompF0_Robert_Ryne-071.pdf}.

\bibitem[Borland et~al.(2021)Borland, Lindberg, and Soliday]{ipac2021_borland}
M.~Borland, R.R. Lindberg, and R.~Soliday.
\newblock Tools for use of generalized gradient expansions in accelerator
  simulations.
\newblock In \emph{Proc. of IPAC}, page MOPAB059, 2021.

\bibitem[Bojt\'ar(2020)]{PhysRevAccelBeams.23.104002}
Lajos Bojt\'ar.
\newblock Frequency analysis and dynamic aperture studies in a low energy
  antiproton ring with realistic 3d magnetic fields.
\newblock \emph{Phys. Rev. Accel. Beams}, 23:\penalty0 104002, 10 2020.
\newblock \doi{10.1103/PhysRevAccelBeams.23.104002}.
\newblock URL
  \url{https://link.aps.org/doi/10.1103/PhysRevAccelBeams.23.104002}.

\bibitem[Manikonda and Berz(2006)]{manikonda2006}
Shashikant Manikonda and Martin Berz.
\newblock Multipole expansion solution of the laplace equation using surface
  data.
\newblock \emph{Nuclear Instruments and Methods in Physics Research Section A:
  Accelerators, Spectrometers, Detectors and Associated Equipment},
  558\penalty0 (1):\penalty0 175--183, 2006.
\newblock ISSN 0168-9002.
\newblock \doi{https://doi.org/10.1016/j.nima.2005.11.081}.
\newblock URL
  \url{https://www.sciencedirect.com/science/article/pii/S0168900205021352}.
\newblock Proceedings of the 8th International Computational Accelerator
  Physics Conference.

\bibitem[Manikonda and Berz(2005)]{manikonda2005}
Shashikant Manikonda and Martin Berz.
\newblock An accurate high-order method to solve the helmholtz boundary value
  problem for the 3d laplace equation.
\newblock \emph{International Journal of Pure and Applied Mathematics},
  23:\penalty0 365--378, 01 2005.

\bibitem[Saldin et~al.(1998)Saldin, Schneidmiller, and Yurkov]{SALDIN1998}
E.L. Saldin, E.A. Schneidmiller, and M.V. Yurkov.
\newblock Analytical treatment of the radiative interaction of electrons in a
  bunch passing a bending magnet.
\newblock \emph{Nuclear Instruments and Methods in Physics Research Section A:
  Accelerators, Spectrometers, Detectors and Associated Equipment},
  407\penalty0 (1):\penalty0 112--115, 1998.
\newblock ISSN 0168-9002.
\newblock \doi{https://doi.org/10.1016/S0168-9002(97)01378-8}.
\newblock URL
  \url{https://www.sciencedirect.com/science/article/pii/S0168900297013788}.

\bibitem[Katz(2015)]{katz}
Daniel~S. Katz.
\newblock Scientific software challenges and community responses, 2015.
\newblock URL
  \url{https://www.slideshare.net/danielskatz/scientific-software-challenges-and-community-responses}.
\newblock Talk given at RTI International on 7 December 2015, discussing 12
  scientific software challenges and how the scientific software community is
  responding to them.

\bibitem[ssi()]{ssi}
URL \url{https://www.software.ac.uk/}.

\bibitem[Sagan et~al.(2020)Sagan, Huebl, Vay, Bruhwiler, Ryne, Ng, and
  Lehe]{LOI_toolkit}
David Sagan, Axel Huebl, Jean-Luc Vay, David Bruhwiler, Robert Ryne, Cho Ng,
  and Rémi Lehe.
\newblock {Beam Dynamics Toolkit}.
\newblock \emph{Snowmass21 LOI}, 2020.
\newblock URL
  \url{https://www.snowmass21.org/docs/files/summaries/CompF/SNOWMASS21-CompF2_CompF0_Sagan-077.pdf}.

\bibitem[Vay et~al.(2020{\natexlab{b}})Vay, Huebl, Sagan, Bruhwiler, Lehe, Ng,
  Liu, Qiang, Ryne, Stern, Friedman, Th\'evenet, Vincenti, Gessner, Cowan,
  Adelmann, Bussmann, and Bastrakov]{LOI_ecosystem}
Jean-Luc Vay, Axel Huebl, David Sagan, David Bruhwiler, Rémi Lehe, Cho-Kuen
  Ng, Ao~Liu, Ji~Qiang, Robert Ryne, Eric Stern, Alex Friedman, Maxence
  Th\'evenet, Henri Vincenti, Spencer Gessner, Benjamin Cowan, Andreas
  Adelmann, Michael Bussmann, and Sergei Bastrakov.
\newblock A modular community ecosystem for multiphysics particle accelerator
  modeling and design.
\newblock \emph{Snowmass21 LOI}, 2020{\natexlab{b}}.
\newblock URL
  \url{https://www.snowmass21.org/docs/files/summaries/CompF/SNOWMASS21-CompF2_CompF0-AF1_AF0_Vay-070.pdf}.

\bibitem[Sagan and Smith(2005)]{tao}
D.~Sagan and J.C. Smith.
\newblock The tao accelerator simulation program.
\newblock In \emph{Proceedings of the 2005 Particle Accelerator Conference},
  pages 4159--4161, 2005.
\newblock \doi{10.1109/PAC.2005.1591750}.

\bibitem[Iselin(1984)]{MAD2005}
Christoph Iselin.
\newblock The mad program.
\newblock In Winfried Busse and Roman Zelazny, editors, \emph{Computing in
  Accelerator Design and Operation}, pages 146--151, Berlin, Heidelberg, 1984.
  Springer Berlin Heidelberg.
\newblock ISBN 978-3-540-39130-2.

\bibitem[Borland(2000)]{elegant}
M.~Borland.
\newblock {elegant: A Flexible SDDS-Compliant Code for Accelerator Simulation}.
\newblock In \emph{{6th International Computational Accelerator Physics
  Conference (ICAP 2000)}}, 2000.
\newblock \doi{10.2172/761286}.

\bibitem[Qiang et~al.(2020)Qiang, Zhang, Vay, Huebl, Grote, Sonnad, and
  Mitchell]{LOI_Poisson}
Ji~Qiang, He~Zhang, Jean-Luc Vay, Axel Huebl, David Grote, Kiran Sonnad, and
  Chad Mitchell.
\newblock {A Parallel Poisson Solver Library for Accelerator Modeling
  Applications}.
\newblock \emph{Snowmass21 LOI}, 2020.
\newblock URL
  \url{https://www.snowmass21.org/docs/files/summaries/CompF/SNOWMASS21-CompF2_CompF0-050.pdf}.

\bibitem[Terebilo(2001)]{at}
A.~Terebilo.
\newblock Accelerator modeling with matlab accelerator toolbox.
\newblock In \emph{PACS2001. Proceedings of the 2001 Particle Accelerator
  Conference (Cat. No.01CH37268)}, volume~4, pages 3203--3205, 2001.

\bibitem[Sagan(2006)]{bmad}
D.~Sagan.
\newblock Bmad: A relativistic charged particle simulation library.
\newblock \emph{Nuclear Instruments and Methods in Physics Research Section A},
  558\penalty0 (1):\penalty0 356 -- 359, 2006.

\bibitem[Berz(1993)]{cosy}
Martin Berz.
\newblock New features in cosy infinity.
\newblock \emph{AIP Conference Proceedings}, 297\penalty0 (1):\penalty0
  267--278, 1993.

\bibitem[Molson et~al.(2010)Molson, Owen, Toader, and Barlow]{merlin}
J.~Molson, H.~Owen, {A. M.} Toader, and {R. J.} Barlow.
\newblock Advances with merlin - a beam tracking code.
\newblock \emph{IPAC 2010 - 1st International Particle Accelerator Conference},
  pages 1853--1855, 2010.

\bibitem[Fiedman et~al.(1992)Fiedman, Grote, Callahan, Langdon, and
  Haber]{warp}
A.~Fiedman, D.~P. Grote, D.~A. Callahan, B.~A. Langdon, and I.~Haber.
\newblock 3d particle simulation of beams using the warp code.
\newblock \emph{Part. Accel.}, 37-38:\penalty0 131--139, 1992.

\bibitem[Qiang et~al.(2000)Qiang, Ryne, Habib, and Decyk]{QIANG2000434}
Ji~Qiang, Robert~D. Ryne, Salman Habib, and Viktor Decyk.
\newblock An object-oriented parallel particle-in-cell code for beam dynamics
  simulation in linear accelerators.
\newblock \emph{Journal of Computational Physics}, 163\penalty0 (2):\penalty0
  434 -- 451, 2000.
\newblock ISSN 0021-9991.
\newblock \doi{https://doi.org/10.1006/jcph.2000.6570}.

\bibitem[Forest et~al.(2006)Forest, Nogiwa, and Schmidt]{ptc}
E.~Forest, Y.~Nogiwa, and F.~Schmidt.
\newblock The fpp and ptc libraries.
\newblock \emph{Proceedings of ICAP 2006}, pages 17--21, 2006.

\bibitem[Mayes et~al.(2021)Mayes, Fuoss, Garrahan, Slepicka, Halavanau,
  Krzywinski, Edelen, F.~Ji, Neveu, Huebl, Lehe, Gupta, Gulliford, Sagan, E,
  and Fortmann-Grote]{lume}
C.~E. Mayes, P.~H. Fuoss, J.~R. Garrahan, H.~Slepicka, A.~Halavanau,
  J.~Krzywinski, A.~L. Edelen, W.~Lou F.~Ji, N.~R. Neveu, A.~Huebl, R.~Lehe,
  L.~Gupta, C.~M. Gulliford, D.~C. Sagan, J.~C. E, and C.~Fortmann-Grote.
\newblock Lightsource unified modeling environment (lume), a start-to-end
  simulation ecosystem.
\newblock In \emph{Proc. of IPAC}, page THPAB217, 2021.

\bibitem[pan()]{panosc}
The {{Photon}} and {{Neutron Open Science Cloud}} ({{PaNOSC}}).
\newblock URL \url{https://www.panosc.eu/}.

\bibitem[sim(2020)]{simex}
{{PaNOSC}}-{{ViNYL}}/{{SimEx}}, May 2020.
\newblock URL \url{https://github.com/PaNOSC-ViNYL/SimEx}.

\bibitem[Dimitrov et~al.(2005)Dimitrov, Bruhwiler, Cary, Messmer, Stoltz,
  Jensen, Feldman, and O'Shea]{1591190}
D.A. Dimitrov, D.L. Bruhwiler, J.R. Cary, P.~Messmer, P.~Stoltz, K.L. Jensen,
  D.W. Feldman, and P.G. O'Shea.
\newblock Development of advanced models for 3d photocathode pic simulations.
\newblock In \emph{Proceedings of the 2005 Particle Accelerator Conference},
  pages 2583--2585, 2005.
\newblock \doi{10.1109/PAC.2005.1591190}.

\bibitem[Nagaitsev et~al.(2021)Nagaitsev, Huang, Power, Vay, Piot, Spentzouris,
  Rosenzweig, Cai, Cousineau, Conde, Hogan, Valishev, Minty, Zolkin, Huang,
  Shiltsev, Seeman, Byrd, Hao, Dunham, Carlsten, Seryi, and
  Patterson]{Nagaitsev2021}
S.~Nagaitsev, Z.~Huang, J.~Power, J.~Vay, P.~Piot, L.~Spentzouris,
  J.~Rosenzweig, Y.~Cai, S.~Cousineau, M.~Conde, M.~Hogan, A.~Valishev,
  M.~Minty, T.~Zolkin, X.~Huang, V.~Shiltsev, J.~Seeman, J.~Byrd, Y.~Hao,
  B.~Dunham, B.~Carlsten, A.~Seryi, and R.~Patterson.
\newblock Accelerator and beam physics research goals and opportunities.
\newblock \emph{arXiv: Accelerator Physics}, 2021.

\bibitem[Zhang et~al.(2017)Zhang, Jin, Sun, Romanus, Bui, Klasky, and
  Parashar]{Zhang2017InmemorySA}
F.~Zhang, Tong Jin, Qian Sun, Melissa Romanus, Hoang Bui, S.~Klasky, and
  M.~Parashar.
\newblock In‐memory staging and data‐centric task placement for coupled
  scientific simulation workflows.
\newblock \emph{Concurrency and Computation: Practice and Experience}, 29,
  2017.

\bibitem[Ayachit et~al.(2016)Ayachit, Bauer, Duque, Eisenhauer, Ferrier, Gu,
  Jansen, Loring, Lukic, Menon, Morozov, O'Leary, Ranjan, Rasquin, Stone,
  Vishwanath, Weber, Whitlock, Wolf, Wu, and Bethel]{7877157}
Utkarsh Ayachit, Andrew Bauer, Earl P.~N. Duque, Greg Eisenhauer, Nicola
  Ferrier, Junmin Gu, Kenneth~E. Jansen, Burlen Loring, Zarija Lukic, Suresh
  Menon, Dmitriy Morozov, Patrick O'Leary, Reetesh Ranjan, Michel Rasquin,
  Christopher~P. Stone, Venkat Vishwanath, Gunther~H. Weber, Brad Whitlock,
  Matthew Wolf, K.~John Wu, and E.~Wes Bethel.
\newblock Performance analysis, design considerations, and applications of
  extreme-scale in situ infrastructures.
\newblock In \emph{SC '16: Proceedings of the International Conference for High
  Performance Computing, Networking, Storage and Analysis}, pages 921--932,
  2016.
\newblock \doi{10.1109/SC.2016.78}.

\bibitem[Larsen et~al.(2017)Larsen, Ahrens, Ayachit, Brugger, Childs, Geveci,
  and Harrison]{alpine2017}
Matthew Larsen, James Ahrens, Utkarsh Ayachit, Eric Brugger, Hank Childs, Berk
  Geveci, and Cyrus Harrison.
\newblock The alpine in situ infrastructure: Ascending from the ashes of
  strawman.
\newblock In \emph{Proceedings of the In Situ Infrastructures on Enabling
  Extreme-Scale Analysis and Visualization}, ISAV'17, page 42–46, New York,
  NY, USA, 2017. Association for Computing Machinery.
\newblock ISBN 9781450351393.
\newblock \doi{10.1145/3144769.3144778}.
\newblock URL \url{https://doi.org/10.1145/3144769.3144778}.

\bibitem[IDE()]{IDEAS}
{IDEAS: Interoperable Design of Extreme-scale Application Software}.
\newblock URL \url{https://ideas-productivity.org}.

\bibitem[xSD()]{xSDK}
{xSDK: Extreme-scale Scientific Software Development Kit}.
\newblock URL \url{http://xsdk.info}.

\bibitem[{The HYPRE Team}()]{HYPRE}
{The HYPRE Team}.
\newblock {HYPRE}.
\newblock URL
  \url{https://computing.llnl.gov/projects/hypre-scalable-linear-solvers-multigrid-methods}.

\bibitem[Hudson et~al.(2020)Hudson, Larson, Wild, Bindel, and
  Navarro]{libEnsemble}
Stephen Hudson, Jeffrey Larson, Stefan~M. Wild, David Bindel, and John-Luke
  Navarro.
\newblock {libEnsemble} users manual.
\newblock Technical Report Revision 0.7.0, Argonne National Laboratory, 2020.
\newblock URL
  \url{https://buildmedia.readthedocs.org/media/pdf/libensemble/latest/libensemble.pdf}.

\bibitem[{The AMReX Team}()]{AMReX}
{The AMReX Team}.
\newblock {AMReX}.
\newblock URL \url{https://amrex-codes.github.io/}.

\bibitem[{The Copa-Cabana Team}()]{CopaCabana}
{The Copa-Cabana Team}.
\newblock {Copa-Cabana}.
\newblock URL \url{https://github.com/ECP-copa/Cabana}.

\bibitem[{The Portage Team}()]{Portage}
{The Portage Team}.
\newblock {Portage}.
\newblock URL \url{https://github.com/laristra/portage/releases}.

\bibitem[REA()]{REANA}
{REANA Reproducible research data analysis platform }.
\newblock URL \url{http://reanahub.io}.

\bibitem[CAM()]{CAMPA}
{CAMPA: Consortium for Advanced Modeling of Particle Accelerators}.
\newblock URL \url{http://campa.lbl.gov}.

\bibitem[CER()]{CERNdata}
{CERN Data Portal}.
\newblock URL \url{http://opendata.cern.ch}.

\bibitem[FAI()]{FAIRdata}
{FAIR Principles}.
\newblock URL \url{https://www.go-fair.org/fair-principles/}.

\bibitem[ATL(2019)]{ATL-PHYS-PUB-2019-032}
{RECAST framework reinterpretation of an ATLAS Dark Matter Search constraining
  a model of a dark Higgs boson decaying to two $b$-quarks}.
\newblock Technical Report ATL-PHYS-PUB-2019-032, CERN, Geneva, 8 2019.
\newblock URL \url{https://cds.cern.ch/record/2686290}.

\bibitem[Godoy et~al.(2020)Godoy, Podhorszki, Wang, Atkins, Eisenhauer, Gu,
  Davis, Choi, Germaschewski, Huck, Huebl, Kim, Kress, Kurc, Liu, Logan, Mehta,
  Ostrouchov, Parashar, Poeschel, Pugmire, Suchyta, Takahashi, Thompson,
  Tsutsumi, Wan, Wolf, Wu, and Klasky]{ADIOS2}
William~F. Godoy, Norbert Podhorszki, Ruonan Wang, Chuck Atkins, Greg
  Eisenhauer, Junmin Gu, Philip Davis, Jong Choi, Kai Germaschewski, Kevin
  Huck, Axel Huebl, Mark Kim, James Kress, Tahsin Kurc, Qing Liu, Jeremy Logan,
  Kshitij Mehta, George Ostrouchov, Manish Parashar, Franz Poeschel, David
  Pugmire, Eric Suchyta, Keichi Takahashi, Nick Thompson, Seiji Tsutsumi,
  Lipeng Wan, Matthew Wolf, Kesheng Wu, and Scott Klasky.
\newblock Adios 2: The adaptable input output system. a framework for
  high-performance data management.
\newblock \emph{SoftwareX}, 12:\penalty0 100561, 2020.
\newblock ISSN 2352-7110.
\newblock \doi{https://doi.org/10.1016/j.softx.2020.100561}.
\newblock URL
  \url{https://www.sciencedirect.com/science/article/pii/S2352711019302560}.

\bibitem[{The HDF Group}()]{HDF5}
{The HDF Group}.
\newblock {Hierarchical data format version 5}.
\newblock URL \url{http://www.hdfgroup.org/HDF5}.

\bibitem[Huebl et~al.(2017)Huebl, Widera, Schmitt, Matthes, Podhorszki, Choi,
  Klasky, and Bussmann]{Huebl2017}
Axel Huebl, Ren{\'e} Widera, Felix Schmitt, Alexander Matthes, Norbert
  Podhorszki, Jong~Youl Choi, Scott Klasky, and Michael Bussmann.
\newblock On the scalability of data reduction techniques in current and
  upcoming hpc systems from an application perspective.
\newblock In Julian~M. Kunkel, Rio Yokota, Michela Taufer, and John Shalf,
  editors, \emph{High Performance Computing}, pages 15--29, Cham, 2017.
  Springer International Publishing.
\newblock ISBN 978-3-319-67630-2.

\bibitem[ope()]{openPMDprojects}
{Curated catalogue of projects supporting openPMD}.
\newblock URL \url{https://github.com/openPMD/openPMD-projects}.

\bibitem[PIC()]{PICMI}
{PICMI: Standard input format for Particle-In-Cell codes}.
\newblock URL \url{https://github.com/picmi-standard}.

\bibitem[Vay et~al.(2021)Vay, Huebl, Cook, England, Niedermayer, Piot, and
  D.]{VayICFA2021}
J.-L. Vay, A.~Huebl, N.M. Cook, R.~J. England, U.~Niedermayer, P.~Piot, and
  Winklehner D.
\newblock {Modeling of Advanced Accelerator Concepts}.
\newblock \emph{This Issue}, 2021.

\bibitem[Friedman et~al.(1992)Friedman, Grote, and Haber]{Friedman:1992}
Alex Friedman, David~P. Grote, and Irving Haber.
\newblock {Three-dimensional particle simulation of heavy-ion fusion beams}.
\newblock \emph{Phys. Fluids B}, 4:\penalty0 2203--2210, 1992.
\newblock \doi{10.1063/1.860024}.

\bibitem[Hockney and Eastwood(1988)]{Hockney1966ComputerSU}
R.~Hockney and J.~Eastwood.
\newblock Computer simulation using particles.
\newblock 1988.

\bibitem[Amundson et~al.(2006)Amundson, Spentzouris, Qiang, and
  Ryne]{AMUNDSON2006229}
J.~Amundson, P.~Spentzouris, J.~Qiang, and R.~Ryne.
\newblock Synergia: An accelerator modeling tool with 3-d space charge.
\newblock \emph{Journal of Computational Physics}, 211\penalty0 (1):\penalty0
  229 -- 248, 2006.
\newblock ISSN 0021-9991.
\newblock \doi{https://doi.org/10.1016/j.jcp.2005.05.024}.

\bibitem[Qiang(2017)]{PhysRevAccelBeams.20.014203}
Ji~Qiang.
\newblock Symplectic multiparticle tracking model for self-consistent
  space-charge simulation.
\newblock \emph{Phys. Rev. Accel. Beams}, 20:\penalty0 014203, 1 2017.
\newblock \doi{10.1103/PhysRevAccelBeams.20.014203}.

\bibitem[Qiang(2018)]{PhysRevAccelBeams.21.054201}
Ji~Qiang.
\newblock Symplectic particle-in-cell model for space-charge beam dynamics
  simulation.
\newblock \emph{Phys. Rev. Accel. Beams}, 21:\penalty0 054201, 5 2018.
\newblock \doi{10.1103/PhysRevAccelBeams.21.054201}.

\bibitem[Greengard and Rokhlin(1987)]{GREENGARD1987325}
L~Greengard and V~Rokhlin.
\newblock A fast algorithm for particle simulations.
\newblock \emph{Journal of Computational Physics}, 73\penalty0 (2):\penalty0
  325--348, 1987.
\newblock ISSN 0021-9991.
\newblock \doi{https://doi.org/10.1016/0021-9991(87)90140-9}.
\newblock URL
  \url{https://www.sciencedirect.com/science/article/pii/0021999187901409}.

\bibitem[Ying et~al.(2003)Ying, Biros, Zorin, and
  Langston]{10.1145/1048935.1050165}
Lexing Ying, George Biros, Denis Zorin, and Harper Langston.
\newblock A new parallel kernel-independent fast multipole method.
\newblock In \emph{Proceedings of the 2003 ACM/IEEE Conference on
  Supercomputing}, SC '03, page~14, New York, NY, USA, 2003. Association for
  Computing Machinery.
\newblock ISBN 1581136951.
\newblock \doi{10.1145/1048935.1050165}.
\newblock URL \url{https://doi.org/10.1145/1048935.1050165}.

\bibitem[Lashuk et~al.(2012)Lashuk, Chandramowlishwaran, Langston, Nguyen,
  Sampath, Shringarpure, Vuduc, Ying, Zorin, and
  Biros]{10.1145/2160718.2160740}
Ilya Lashuk, Aparna Chandramowlishwaran, Harper Langston, Tuan-Anh Nguyen,
  Rahul Sampath, Aashay Shringarpure, Richard Vuduc, Lexing Ying, Denis Zorin,
  and George Biros.
\newblock A massively parallel adaptive fast multipole method on heterogeneous
  architectures.
\newblock \emph{Commun. ACM}, 55\penalty0 (5):\penalty0 101–109, May 2012.
\newblock ISSN 0001-0782.
\newblock \doi{10.1145/2160718.2160740}.
\newblock URL \url{https://doi.org/10.1145/2160718.2160740}.

\bibitem[Zhang and Berz(2011)]{ZHANG2011338}
He~Zhang and Martin Berz.
\newblock The fast multipole method in the differential algebra framework.
\newblock \emph{Nuclear Instruments and Methods in Physics Research Section A:
  Accelerators, Spectrometers, Detectors and Associated Equipment},
  645\penalty0 (1):\penalty0 338 -- 344, 2011.
\newblock ISSN 0168-9002.
\newblock \doi{https://doi.org/10.1016/j.nima.2011.01.053}.
\newblock The Eighth International Conference on Charged Particle Optics.

\bibitem[Langston et~al.(2011)Langston, Greengard, and Zorin]{LGZ1-2011}
M.~Harper Langston, Leslie Greengard, and D.~Zorin.
\newblock A free-space adaptive {FMM}-based {PDE} solver in three dimensions.
\newblock \emph{Communications in Applied Mathematics and Computational
  Science}, 6\penalty0 (1):\penalty0 79--122, 2011.
\newblock \doi{10.2140/camcos.2011.6.79}.

\bibitem[Biros et~al.(2004)Biros, Ying, and Zorin]{BIROS2004317}
George Biros, Lexing Ying, and Denis Zorin.
\newblock A fast solver for the stokes equations with distributed forces in
  complex geometries.
\newblock \emph{Journal of Computational Physics}, 193\penalty0 (1):\penalty0
  317 -- 348, 2004.
\newblock ISSN 0021-9991.
\newblock \doi{https://doi.org/10.1016/j.jcp.2003.08.011}.

\bibitem[Klockner et~al.(2013)Klockner, Barnett, Greengard, and
  O'Neil]{KLOCKNER2013332}
Andreas Klockner, Alexander Barnett, Leslie Greengard, and Michael O'Neil.
\newblock Quadrature by expansion: A new method for the evaluation of layer
  potentials.
\newblock \emph{Journal of Computational Physics}, 252:\penalty0 332--349,
  2013.
\newblock ISSN 0021-9991.
\newblock \doi{https://doi.org/10.1016/j.jcp.2013.06.027}.
\newblock URL
  \url{https://www.sciencedirect.com/science/article/pii/S0021999113004579}.

\bibitem[Morse et~al.(2021)Morse, Rahimian, and Zorin]{MORSE2021110511}
Matthew~J. Morse, Abtin Rahimian, and Denis Zorin.
\newblock A robust solver for elliptic pdes in 3d complex geometries.
\newblock \emph{Journal of Computational Physics}, 442:\penalty0 110511, 2021.
\newblock ISSN 0021-9991.
\newblock \doi{https://doi.org/10.1016/j.jcp.2021.110511}.
\newblock URL
  \url{https://www.sciencedirect.com/science/article/pii/S002199912100406X}.

\bibitem[Vico et~al.(2016)Vico, Greengard, and
  Ferrando]{10.1016/j.jcp.2016.07.028}
Felipe Vico, Leslie Greengard, and Miguel Ferrando.
\newblock Fast convolution with free-space green's functions.
\newblock \emph{J. Comput. Phys.}, 323\penalty0 (C):\penalty0 191–203,
  October 2016.
\newblock ISSN 0021-9991.
\newblock \doi{10.1016/j.jcp.2016.07.028}.
\newblock URL \url{https://doi.org/10.1016/j.jcp.2016.07.028}.

\bibitem[Zou et~al.(2021)Zou, Kim, and Cerfon]{zou2021fftbased}
Junyi Zou, Eugenia Kim, and Antoine~J. Cerfon.
\newblock Fft-based free space poisson solvers: why vico-greengard-ferrando
  should replace hockney-eastwood, 2021.

\bibitem[Behrens et~al.(2014)Behrens, Decker, Ding, Dolgashev, Frisch, Huang,
  Krejcik, Loos, Lutman, Maxwell, et~al.]{behrens2014few}
C~Behrens, F-J Decker, Y~Ding, VA~Dolgashev, J~Frisch, Z~Huang, P~Krejcik,
  H~Loos, A~Lutman, TJ~Maxwell, et~al.
\newblock Few-femtosecond time-resolved measurements of x-ray free-electron
  lasers.
\newblock \emph{Nature communications}, 5\penalty0 (1):\penalty0 1--7, 2014.
\newblock \doi{https://doi.org/10.1038/ncomms4762}.

\bibitem[Joshi et~al.(2018)Joshi, Adli, An, Clayton, Corde, Gessner, Hogan,
  Litos, Lu, Marsh, et~al.]{joshi2018plasma}
C~Joshi, Erik Adli, W~An, CE~Clayton, Sebastian Corde, S~Gessner, MJ~Hogan,
  M~Litos, W~Lu, KA~Marsh, et~al.
\newblock Plasma wakefield acceleration experiments at facet ii.
\newblock \emph{Plasma Physics and Controlled Fusion}, 60\penalty0
  (3):\penalty0 034001, 2018.
\newblock \doi{https://doi.org/10.1088/1361-6587/aaa2e3}.

\bibitem[Malyzhenkov et~al.(2020)Malyzhenkov, Arbelo, Craievich, Dijkstal,
  Ferrari, Reiche, Schietinger, Jurani{\'c}, and Prat]{malyzhenkov2020single}
Alexander Malyzhenkov, Yunieski~P Arbelo, Paolo Craievich, Philipp Dijkstal,
  Eugenio Ferrari, Sven Reiche, Thomas Schietinger, Pavle Jurani{\'c}, and
  Eduard Prat.
\newblock Single-and two-color attosecond hard x-ray free-electron laser pulses
  with nonlinear compression.
\newblock \emph{Physical Review Research}, 2\penalty0 (4):\penalty0 042018,
  2020.
\newblock \doi{https://doi.org/10.1103/PhysRevResearch.2.042018}.

\bibitem[Aihara et~al.(2019)Aihara, Bagger, Bambade, Barish, Behnke, Bellerive,
  Berggren, Brau, Breidenbach, Bozovic-Jelisavcic,
  et~al.]{aihara2019international}
Hiroaki Aihara, Jonathan Bagger, Philip Bambade, Barry Barish, Ties Behnke,
  Alain Bellerive, Mikael Berggren, James Brau, Martin Breidenbach, Ivanka
  Bozovic-Jelisavcic, et~al.
\newblock the international linear collider. a global project.
\newblock \emph{arXiv preprint arXiv:1901.09829}, 2019.
\newblock URL \url{https://arxiv.org/pdf/1903.01629.pdf}.

\bibitem[Bane and Emma(2005)]{bane2005litrack}
KLF Bane and P~Emma.
\newblock Litrack: a fast longitudinal phase space tracking code with graphical
  user interface.
\newblock In \emph{Proceedings of the 2005 Particle Accelerator Conference},
  pages 4266--4268. IEEE, 2005.
\newblock \doi{https://doi.org/10.1109/PAC.2005.1591786}.

\bibitem[Scheinker and Gessner(2015)]{scheinker2015adaptive}
Alexander Scheinker and Spencer Gessner.
\newblock Adaptive method for electron bunch profile prediction.
\newblock \emph{Physical Review Special Topics-Accelerators and Beams},
  18\penalty0 (10):\penalty0 102801, 2015.

\bibitem[Scheinker et~al.(2021)Scheinker, Cropp, Paiagua, and
  Filippetto]{scheinker2021adaptive}
Alexander Scheinker, Frederick Cropp, Sergio Paiagua, and Daniele Filippetto.
\newblock Adaptive deep learning for time-varying systems with hidden
  parameters: Predicting changing input beam distributions of compact particle
  accelerators.
\newblock \emph{arXiv preprint arXiv:2102.10510}, 2021.
\newblock URL \url{https://arxiv.org/ftp/arxiv/papers/2102/2102.10510.pdf}.

\bibitem[Arute et~al.(2019)Arute, Arya, Babbush, Bacon, Bardin, Barends,
  Biswas, Boixo, Brandao, Buell, et~al.]{arute2019quantum}
Frank Arute, Kunal Arya, Ryan Babbush, Dave Bacon, Joseph~C Bardin, Rami
  Barends, Rupak Biswas, Sergio Boixo, Fernando~GSL Brandao, David~A Buell,
  et~al.
\newblock Quantum supremacy using a programmable superconducting processor.
\newblock \emph{Nature}, 574\penalty0 (7779):\penalty0 505--510, 2019.

\bibitem[Zhong et~al.(2020)Zhong, Wang, Deng, Chen, Peng, Luo, Qin, Wu, Ding,
  Hu, et~al.]{zhong2020quantum}
Han-Sen Zhong, Hui Wang, Yu-Hao Deng, Ming-Cheng Chen, Li-Chao Peng, Yi-Han
  Luo, Jian Qin, Dian Wu, Xing Ding, Yi~Hu, et~al.
\newblock Quantum computational advantage using photons.
\newblock \emph{Science}, 370\penalty0 (6523):\penalty0 1460--1463, 2020.

\bibitem[IBM(2020)]{IBMQexperience}
Ibm quantum experience.
\newblock \url{https://www.ibm.com/quantum-computing/technology/experience/},
  2020.

\bibitem[DWa(2020)]{DWaveLeap2}
Leap$^2$.
\newblock \url{https://www.dwavesys.com/}, 2020.

\bibitem[Ama(2020)]{AmazonQSulution}
Amazon quantum solutions lab.
\newblock \url{https://aws.amazon.com/quantum-solutions-lab/}, 2020.

\bibitem[Azu(2020)]{AzureQ}
Azure quantum.
\newblock \url{https://azure.microsoft.com/en-us/services/quantum/}, 2020.

\bibitem[Harrow et~al.(2009)Harrow, Hassidim, and Lloyd]{harrow2009quantum}
Aram~W Harrow, Avinatan Hassidim, and Seth Lloyd.
\newblock Quantum algorithm for linear systems of equations.
\newblock \emph{Physical review letters}, 103\penalty0 (15):\penalty0 150502,
  2009.

\bibitem[Clader et~al.(2013)Clader, Jacobs, and
  Sprouse]{clader2013preconditioned}
B~David Clader, Bryan~C Jacobs, and Chad~R Sprouse.
\newblock Preconditioned quantum linear system algorithm.
\newblock \emph{Physical review letters}, 110\penalty0 (25):\penalty0 250504,
  2013.

\bibitem[Childs et~al.(2017)Childs, Kothari, and Somma]{childs2017quantum}
Andrew~M Childs, Robin Kothari, and Rolando~D Somma.
\newblock Quantum algorithm for systems of linear equations with exponentially
  improved dependence on precision.
\newblock \emph{SIAM Journal on Computing}, 46\penalty0 (6):\penalty0
  1920--1950, 2017.

\bibitem[Bravo-Prieto et~al.(2019)Bravo-Prieto, LaRose, Cerezo, Subasi, Cincio,
  and Coles]{bravo2019variational}
Carlos Bravo-Prieto, Ryan LaRose, Marco Cerezo, Yigit Subasi, Lukasz Cincio,
  and Patrick~J Coles.
\newblock Variational quantum linear solver: A hybrid algorithm for linear
  systems.
\newblock \emph{arXiv preprint arXiv:1909.05820}, 2019.

\bibitem[Lee et~al.(2019)Lee, Joo, and Lee]{lee2019hybrid}
Yonghae Lee, Jaewoo Joo, and Soojoon Lee.
\newblock Hybrid quantum linear equation algorithm and its experimental test on
  ibm quantum experience.
\newblock \emph{Scientific reports}, 9\penalty0 (1):\penalty0 4778, 2019.

\bibitem[Leyton and Osborne(2008)]{leyton2008quantum}
Sarah~K Leyton and Tobias~J Osborne.
\newblock A quantum algorithm to solve nonlinear differential equations.
\newblock \emph{arXiv preprint arXiv:0812.4423}, 2008.

\bibitem[Berry(2014)]{berry2014high}
Dominic~W Berry.
\newblock High-order quantum algorithm for solving linear differential
  equations.
\newblock \emph{Journal of Physics A: Mathematical and Theoretical},
  47\penalty0 (10):\penalty0 105301, 2014.

\bibitem[Arrazola et~al.(2019)Arrazola, Kalajdzievski, Weedbrook, and
  Lloyd]{arrazola2019quantum}
Juan~Miguel Arrazola, Timjan Kalajdzievski, Christian Weedbrook, and Seth
  Lloyd.
\newblock Quantum algorithm for nonhomogeneous linear partial differential
  equations.
\newblock \emph{Physical Review A}, 100\penalty0 (3):\penalty0 032306, 2019.

\bibitem[Childs et~al.(2020)Childs, Liu, and Ostrander]{childs2020high}
Andrew~M Childs, Jin-Peng Liu, and Aaron Ostrander.
\newblock High-precision quantum algorithms for partial differential equations.
\newblock \emph{arXiv preprint arXiv:2002.07868}, 2020.

\bibitem[Xin et~al.(2020)Xin, Wei, Cui, Xiao, Arrazola, Lamata, Kong, Lu,
  Solano, and Long]{xin2020quantum}
Tao Xin, Shijie Wei, Jianlian Cui, Junxiang Xiao, I{\~n}igo Arrazola, Lucas
  Lamata, Xiangyu Kong, Dawei Lu, Enrique Solano, and Guilu Long.
\newblock Quantum algorithm for solving linear differential equations: Theory
  and experiment.
\newblock \emph{Physical Review A}, 101\penalty0 (3):\penalty0 032307, 2020.

\bibitem[Cao et~al.(2013)Cao, Papageorgiou, Petras, Traub, and
  Kais]{cao2013quantum}
Yudong Cao, Anargyros Papageorgiou, Iasonas Petras, Joseph Traub, and Sabre
  Kais.
\newblock Quantum algorithm and circuit design solving the poisson equation.
\newblock \emph{New Journal of Physics}, 15\penalty0 (1):\penalty0 013021,
  2013.

\bibitem[Wang et~al.(2019)Wang, Wang, Li, Fan, Wei, and Gu]{wang2019quantum}
Shengbin Wang, Zhimin Wang, Wendong Li, Lixin Fan, Zhiqiang Wei, and Yongjian
  Gu.
\newblock Quantum fast poisson solver: the algorithm and modular circuit
  design.
\newblock \emph{arXiv preprint arXiv:1910.09756}, 2019.

\bibitem[Engel et~al.(2019)Engel, Smith, and Parker]{engel2019quantum}
Alexander Engel, Graeme Smith, and Scott~E Parker.
\newblock Quantum algorithm for the vlasov equation.
\newblock \emph{Physical Review A}, 100\penalty0 (6):\penalty0 062315, 2019.

\bibitem[Mic(2020)]{MicrosoftQDev}
Q\# development kit.
\newblock \url{https://www.microsoft.com/en-us/quantum/development-kit}, 2020.

\bibitem[Dodin and Startsev(2020)]{dodin2020applications}
Ilya~Y Dodin and Edward~A Startsev.
\newblock On applications of quantum computing to plasma simulations.
\newblock \emph{arXiv preprint arXiv:2005.14369}, 2020.

\bibitem[Vay et~al.(2020{\natexlab{c}})Vay, Bruhwiler, Sagan, Huebl, Lehe, Ng,
  Qiang, Ryne, Thevenet, Vincenti, and Thomas]{LOI_center}
Jean-Luc Vay, David Bruhwiler, David Sagan, Axel Huebl, Remi Lehe, Cho-Kuen Ng,
  Ji~Qiang, Robert Ryne, Maxence Thevenet, Henri Vincenti, and Alex Thomas.
\newblock {Center(s) for Accelerator and Beam Physics Modeling}.
\newblock \emph{Snowmass21 LOI}, 2020{\natexlab{c}}.
\newblock URL
  \url{https://www.snowmass21.org/docs/files/summaries/CompF/SNOWMASS21-CompF2_CompF0-AF1_AF0_Vay-069.pdf}.

\bibitem[sch()]{schwarzman}
{MIT} schwarzman college of computing.
\newblock URL \url{https://computing.mit.edu/}.

\bibitem[qui()]{quics}
{QuICS}.
\newblock URL \url{https://quics.umd.edu/}.

\bibitem[Fisher()]{ibm:quantum}
Chris Fisher.
\newblock {IBM} {\textbar} quantum computing.
\newblock URL \url{https://www.ibm.com/quantum-computing//}.

\bibitem[exa()]{exascale}
Exascale computing project.
\newblock URL \url{https://www.exascaleproject.org/}.

\bibitem[sci()]{scidac}
Scientific discovery through advanced computing.
\newblock URL \url{https://www.scidac.gov/}.

\bibitem[Lehe et~al.(2020{\natexlab{b}})Lehe, Hanuka, Edelen, Huang, Mayes,
  Cook, Huebl, Emma, Djordjevi´c, Kemp, Wilks, Roussel, Thevenet, Vay, Qiang,
  Alves, Tennant, Adelmann, Vincenti, Filippetto, Gessner, Scheinker, Thomas,
  Winklehner, Holt, Yoffe, Jaroszynski, Hoffmann, Debus, and Bussmann.]{LOI_ML}
Remi Lehe, Adi Hanuka, Auralee Edelen, Xiaobiao Huang, Christopher Mayes,
  Nathan Cook, Axel Huebl, Claudio Emma, Blagoje Djordjevi´c, Andreas Kemp,
  Scott Wilks, Ryan Roussel, Maxence Thevenet, Jean-Luc Vay, Ji~Qiang, E.~Paulo
  Alves, Chris Tennant, Andreas Adelmann, Henri Vincenti, Daniele Filippetto,
  Spencer Gessner, Alexander Scheinker, Alec Thomas, Daniel Winklehner,
  George~K. Holt, Samuel~R. Yoffe, Dino~A. Jaroszynski, Nico Hoffmann,
  Alexander Debus, and Michael Bussmann.
\newblock {Machine learning and surrogates models for simulation-based
  optimization of accelerator design}.
\newblock \emph{Snowmass21 LOI}, 2020{\natexlab{b}}.
\newblock URL
  \url{snowmass21.org/docs/files/summaries/CompF/SNOWMASS21-CompF2_CompF3-AF1_AF6_Lehe-075.pdf}.

\bibitem[Huebl et~al.(2020{\natexlab{b}})Huebl, Vay, Lehe, Mayes, Tsai,
  Friedman, Th´evenet, Vincenti, Bruhwiler, Sauers, Cook, Gessner, Bussmann,
  Grote, Schmitz, Cros, Sagan, Yoffe, Edelen, Pousa, and Stark]{LOI_open}
A.~Huebl, J.-L. Vay, R.~Lehe, C.~Mayes, Y.-D. Tsai, A.~Friedman, M.~Th´evenet,
  H.~Vincenti, D.~L. Bruhwiler, A.~Sauers, N.~M. Cook, S.~J. Gessner,
  M.~Bussmann, D.~P. Grote, R.~Schmitz, B.~Cros, D.~Sagan, S.~R. Yoffe, A.~L.
  Edelen, A.~Ferran Pousa, and G.~Stark.
\newblock {Aspiration for Open Science in Accelerator \& Beam Physics Modeling
  }.
\newblock \emph{Snowmass21 LOI}, 2020{\natexlab{b}}.
\newblock URL
  \url{https://www.snowmass21.org/docs/files/summaries/CompF/SNOWMASS21-CompF2_CompF7-AF1_AF0_Huebl-081.pdf}.

\bibitem[Riordan et~al.(2015)Riordan, Hoddeson, and Kolb]{tunnelvisions}
Michael Riordan, Lillian Hoddeson, and Adrienne~W. Kolb.
\newblock \emph{Tunnel Visions: The Rise and Fall of the Superconducting Super
  Collider}.
\newblock University of Chicago Press, 2015.

\bibitem[Ritson(1990{\natexlab{a}})]{ritson1990}
D.~Ritson.
\newblock Slac memorandum: D. ritson to r. schwitters, march 8, 1990.,
  1990{\natexlab{a}}.

\bibitem[Ritson(1990{\natexlab{b}})]{ssctrk}
D.~Ritson.
\newblock {SSCTRK: A particle tracking code for the SSC}, 1990{\natexlab{b}}.
\newblock URL
  \url{https://inis.iaea.org/search/search.aspx?orig_q=RN:21094597}.

\end{thebibliography}

\end{document}